\documentclass[a4paper,11pt]{article}
\usepackage{pos}

\title{Cosmic-Ray Physics in the PeV to EeV Energy Range with the IceCube-Gen2 Surface Array}
\ShortTitle{Cosmic-Ray Physics with IceCube-Gen2}

\author{The IceCube-Gen2 Collaboration \\{\normalsize \normalfont(a complete list of authors can be found at the end of the proceedings)}\\}

\emailAdd{fgs@udel.edu}

\abstract{

IceCube-Gen2 is a proposed neutrino observatory at the South Pole that will build on the success of IceCube and will also serve as a unique detector for cosmic-ray air showers. 
Analogous to the IceTop surface array over IceCube’s deep optical detector, IceCube-Gen2 will also feature a surface array above the optical array deep in the ice. 
As improvement over IceTop, the IceCube-Gen2 surface array will be comprised of elevated detectors to avoid snow coverage, and will combine two types of detectors: scintillation panels that measure air-shower particles on ground and enable a low detection threshold, which is important to serve as a veto for selecting downgoing neutrino candidates, and radio antennas which increase the measurement accuracy for air showers by providing a calorimetric measurement of the electromagnetic shower component and its depth of maximum, $X_\mathrm{max}$. 
As another major advantage, the eight times larger surface area combined with a larger field of view will provide a 30-fold increase for the aperture of surface-deep coincident events. 
With these improvements in statistics and measurement accuracy, IceCube-Gen2 will thus make unique contributions to the particle physics and astrophysics of Galactic cosmic rays in the PeV to EeV energy range, including the search for PeV photon sources. 
This proceeding summarizes the technical design and science case enabled by the IceCube-Gen2 Surface Array.

\vspace{4mm}

{\bfseries Corresponding authors:}
Frank G.~Schroeder$^{1,2*}$\\
{$^{1}$ \itshape Bartol Research Institute, Department of Physics and Astronomy, University of Delaware}\\
{$^{2}$ \itshape Institute for Astroparticle Physics, Karlsruhe Institute of Technology (KIT)}\\[4mm]
$^*$ Presenter
}

\ConferenceLogo{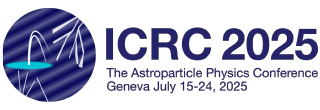}

\FullConference{39th International Cosmic Ray Conference (ICRC2025)\\
 15–24 July 2025\\
Geneva, Switzerland\\}

\begin{document}
\maketitle

\section{Introduction}
IceCube-Gen2 will be a next generation multi-messenger observatory at the South Pole and will extend the capabilities of the existing IceCube observatory for all types of primary particles.
In addition to its main mission of high-energy neutrino astronomy, IceCube-Gen2 will also have excellent capabilities for cosmic rays up to the EeV energy range and will be able to detect PeV gamma rays.
This is due to the unique combination of a surface array measuring extensive air showers in coincidence with the optical array deep in the Antarctic ice.
Compared to IceCube, not only the size of the arrays will be increased by an order of magnitude, but also the measurements will be more accurate thanks to a technical design making use of progress in detection technology~\cite{IceCubeGen2TDR}. 
With the larger arrays, also the zenith angle coverage for surface-deep coincident cosmic-ray events will approximately double which, together with the increased surface area, will lead to an overall 30-fold increase for these unique hybrid events.

This proceeding summarizes the reference technical design and will highlight the cosmic-ray and gamma-ray science cases enabled by the drastic increase in statistics and the new detectors, in particular, the combination of particle and radio detection of air showers.

\begin{figure}[b]
\begin{center}
    \includegraphics[width=0.99\linewidth]{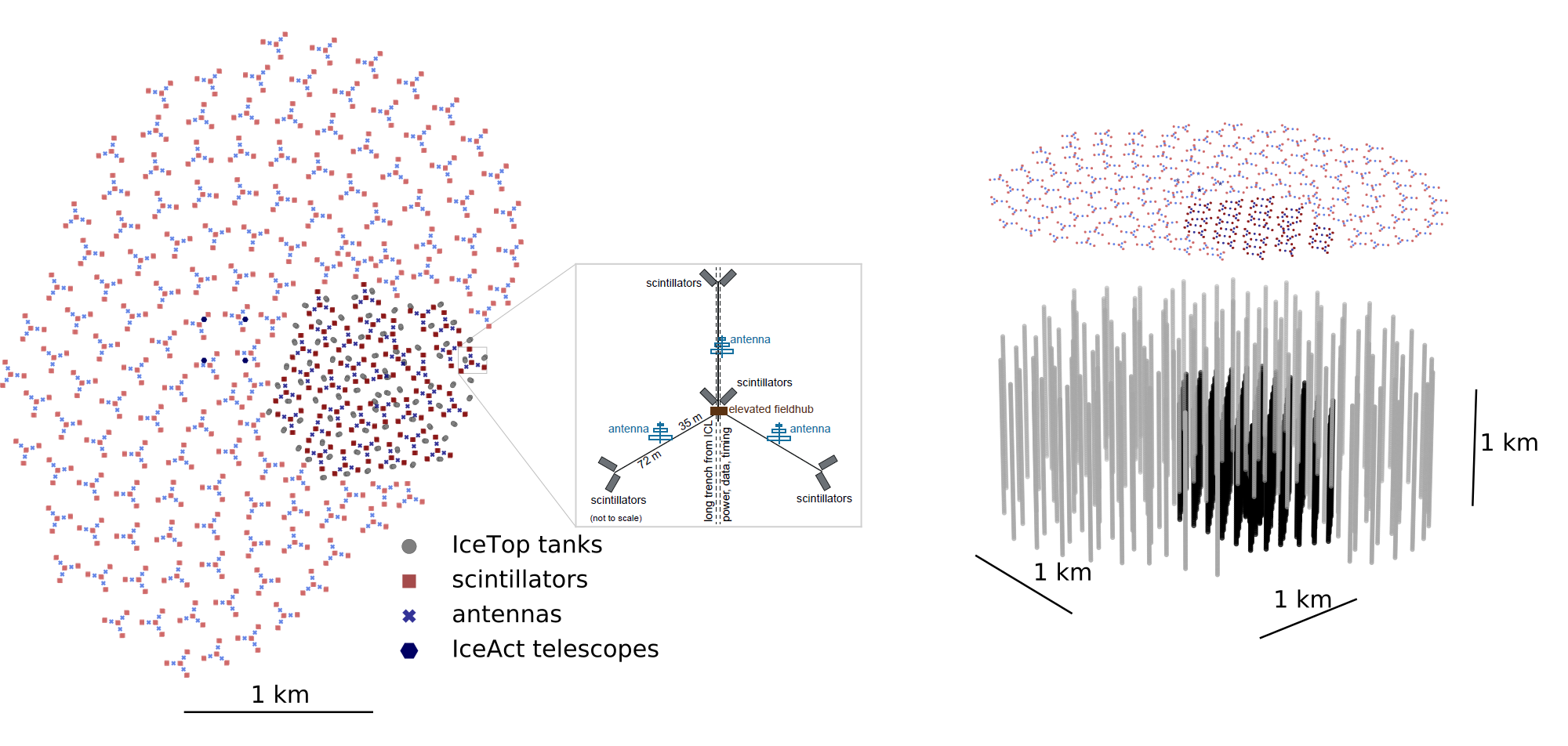}
    \caption{Layout of the IceCube-Gen2 Surface Array: top view on the left-hand side, and side view on the right-hand side, which includes the optical array in the ice underneath the surface array. The denser part shows the existing digital optical modules of IceCube's in-ice array, and the gray dots are existing IceTop detectors. The middle inset shows the layout of a single surface-array station comprised of four pairs of scintillation panels and three radio antennas. The central fieldhub of a station houses the electronics for both the surface station and the string of the corresponding optical array.
    }
    \label{fig_SurfaceLayout}
\end{center}
\end{figure}

\section{Reference Design}
IceCube-Gen2 will be comprised of three arrays: an optical array deep in the ice extending IceCube's optical array, a surface array of elevated scintillation panels and radio antennas covering the footprint of the optical array, and a radio array comprised of antennas in the firn to detect ultra-high-energy neutrinos.

All three arrays contribute to the main mission of neutrino astronomy, e.g., the surface array increases IceCube-Gen2's sensitivity to downgoing neutrinos by providing a veto. 
Although all three arrays will measure cosmic rays and increase IceCube's statistics, the cosmic-ray science case builds particularly on air showers whose shower axis intersects both the surface and optical arrays.
In that case, the optical array measures the high-energy muon content (for $E_\mu \gtrsim 300\,$GeV), and the surface array measures the secondary particles on ground and, at higher energies of $E_\mathrm{shower} \gtrsim 10^{16.5}\,$eV, also the radio emission from the electromagnetic shower component.

For the cosmic-ray science case, the quality of the air-shower measurement at the surface matters while, for the veto purpose, mostly a low detection threshold and a high efficiency are important.
Both, requirements have been taking into account in the reference design of the surface array:
radio antennas provide a calorimetric measurement of the electromagnetic shower energy and measure the depth of shower maximum, $X_\mathrm{max}$~\cite{Huege:2016veh,Schroder:2016hrv}; these are triggered by scintillation panels, which provide a low detection threshold measuring the electromagnetic particles and muons of the air shower and, thus, provide a measure of the shower size at ground.

\begin{figure}[t]
\begin{center}
    \includegraphics[height=6cm]{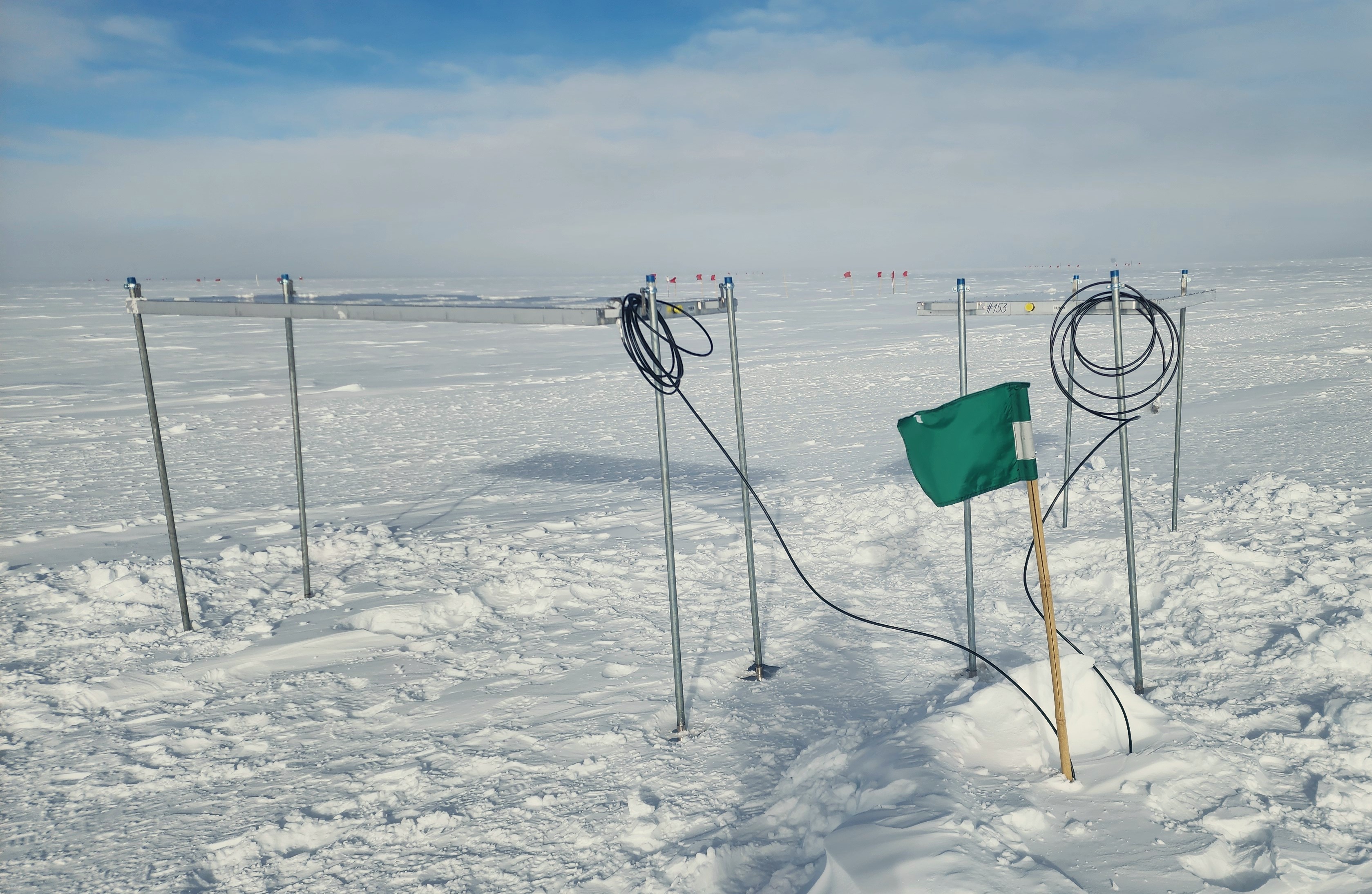}
    \hfill
    \includegraphics[height=6cm]{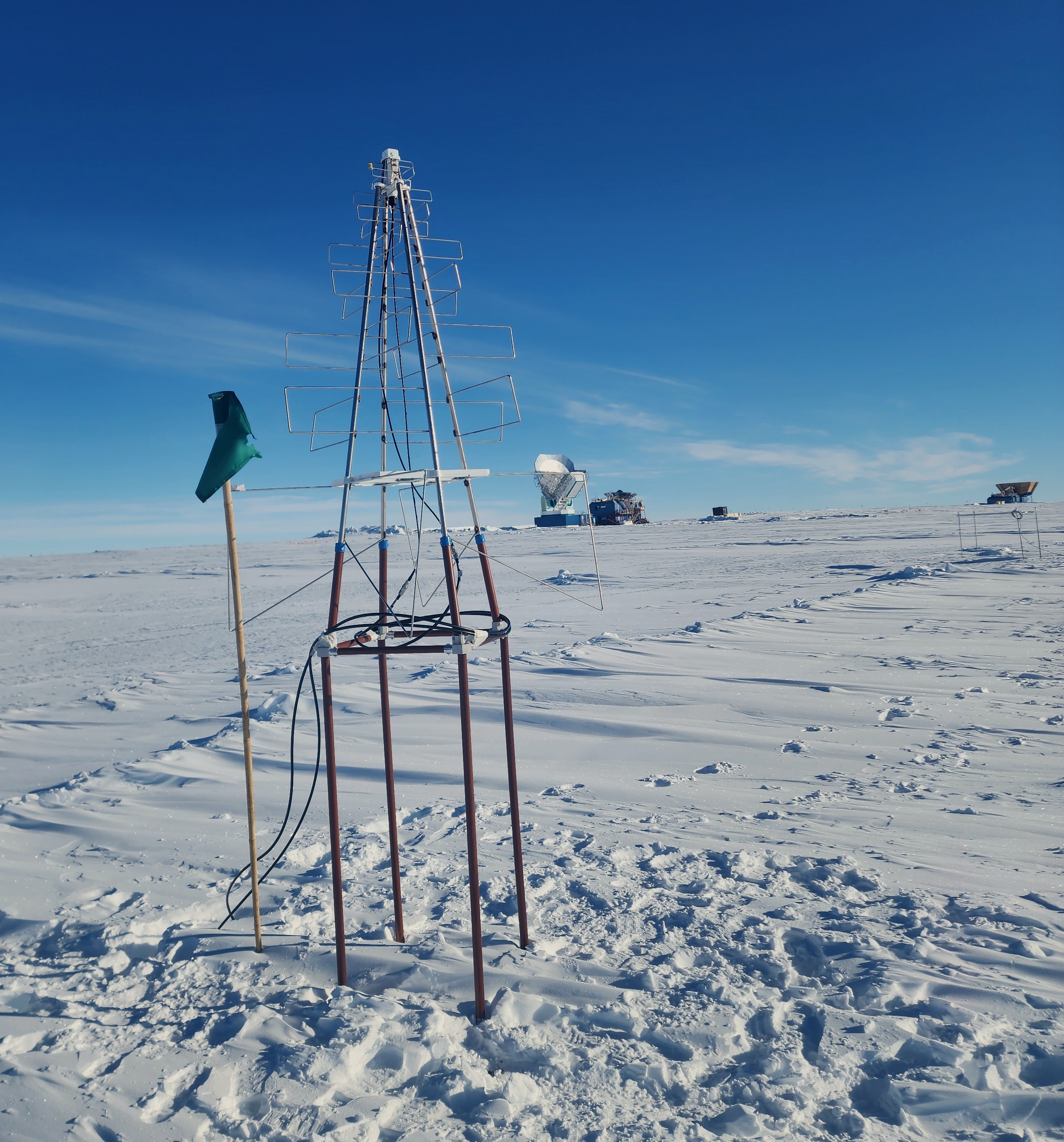}
    \caption{Photos of surface detectors of the IceTop surface enhancement taken in January 2025. The IceCube-Gen2 surface array will be built with the same type of scintillation panels and SKALA-type radio antennas.}
    \label{fig_detectors}
\end{center}
\end{figure}

With an area of about $6\,$km$^2$, the IceCube-Gen2 surface array will be about eight times larger than IceCube's existing surface array, IceTop, which consists of 81 pairs of ice-Cherenkov detectors~\cite{IceCube:2012nn}.
Above each of the 120 strings of optical modules of IceCube-Gen2's optical array, one surface station of 4 pairs of scintillators and 3 SKALA v2~\cite{7297231} radio antennas operating at $70-350\,$MHz will be connected to the same fieldhub housing also the data-acquisition of the corresponding string.
In addition, IceTop is planned to be enhanced by the same type of surface stations~\cite{Shefali:2025icrc,Megha:2025icrc}, such that the full surface area will feature the same type of scintillation panels and radio antennas with approximately $160$ surface stations total (see Fig.~\ref{fig_detectors}).

\begin{figure}[t]
\begin{center}
    \includegraphics[width=0.7\linewidth]{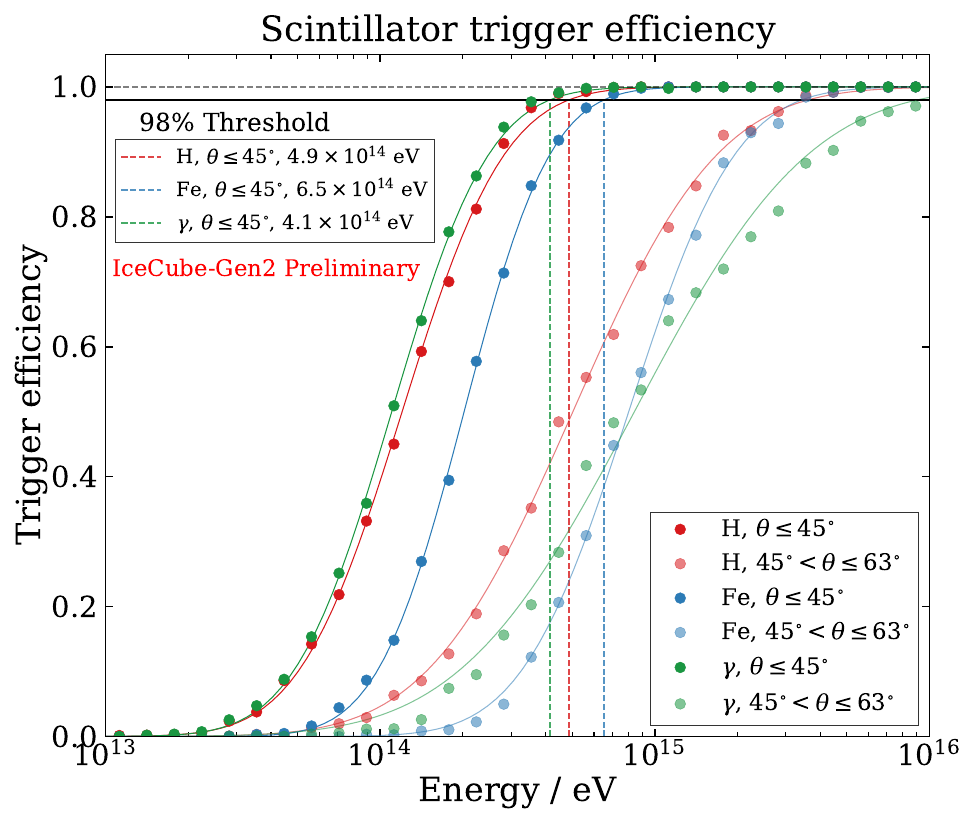}
    \caption{Simulated detection efficiency of the IceCube-Gen2 surface array for different primary particles. The full efficiency threshold for protons and photons is reached around $0.5\,$PeV for modestly inclined showers with zenith angles $\theta < 45^\circ$, and at higher energy for more included showers. The calculation assumes that air showers simulated with CORSIKA are exceed the signal threshold corresponding to half of that of a minimum ionizing particle in at least five scintillation panels (see Ref.~\cite{IceCubeGen2TDR} for details).
    }
    \label{fig_threshold}
\end{center}
\end{figure}

By using elevated detectors easily raisable every few years, the IceCube-Gen2 surface array will improve on one of IceTop's disadvantages.
Continuous snow accumulation above IceTop has increased its detection threshold as the snow absorbs electrons and photons of air showers. 
Several years of operation of a prototype station at the South Pole~\cite{Shefali:2025icrc,Megha:2025icrc} have demonstrated that no snow accumulates on the elevated scintillation panels and radio antennas (see Fig.~\ref{fig_detectors}), so the IceCube-Gen2 surface array will continuously maintain its sub-PeV threshold, reaching full efficiency around $0.5\,$PeV for vertical and mildly inclined protons, and at several PeV for more inclined cosmic rays (Fig.~\ref{fig_threshold}).
At the same time, IceTop will eventually become a detector measuring primarily muons, which will further enhance the science capabilities, as the combined measurement of the sizes of the electromagnetic and muonic shower components and $X_\mathrm{max}$ promises better sensitivity to the mass of the primary cosmic-ray particles~\cite{Holt:2019fnj,Flaggs:2023exc}.

\begin{figure}[t]
\begin{center}
    \includegraphics[width=0.8\linewidth]{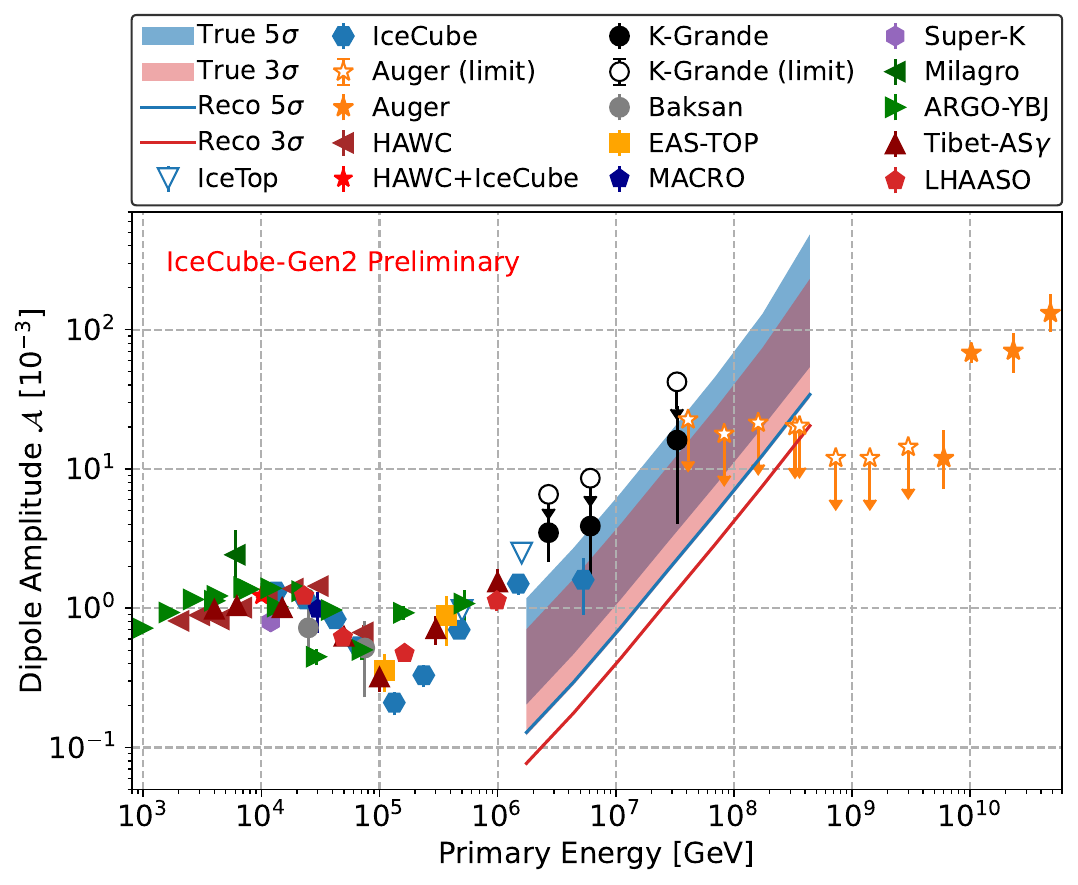}
    \caption{Sensitivity of the IceCube-Gen2 surface array to the large-scale anisotropy of cosmic rays with 10 years of exposure. Shown are the $3\sigma$ and $5\sigma$ sensitivities to the reconstructed dipole of the anisotropy; the bands show the amplitude range of a corresponding true dipole scanning in $10^ \circ$ steps over a declination range of $-80^\circ$ to $-80^\circ$~\cite{Hou:2024edc}. For comparison, measurements and limits of other experiments are shown~\cite{Abeysekara:2018uu,Chiavassa:2015kit,Aglietta:2009INAF,Ambrosio:2003un,Guillian:2005ko,Abdo:2009nasw,Bartoli:2015nu,Amenomori:2005hk,Aartsen:2013au,IceCube:2024pnx,PierreAuger:2024fgl,Gao:2021su}. All points have been calibrated from partial‐sky to full‐sky coverage using a geometric factor that accounts for each experiment’s field of view~\cite{IceCube-Gen2:2023dxt}.}
    \label{fig_anisotropy}
\end{center}
\end{figure}

\section{Cosmic-Ray Physics with IceCube-Gen2}
The higher measurement quality of air showers in combination with the increase in statistics will enable cosmic-ray science with IceCube-Gen2 beyond the current capabilities of IceCube.
In particular in the energy range of the presumed Galactic-to-extragalactic transition of approximately $10^{16.5}-10^{18.5}\,$eV, the combination of the scintillation panels, radio antennas, and in-ice array promises unprecedented precision for the individual cosmic-ray air showers.

This precision can be used to study hadronic interaction in air showers, e.g., by scrutinizing hadronic interaction models. 
Until now, only average parameters, such as the mean muon density over energy have been tested against model predictions~\cite{IceCubeCollaboration:2022tla,Verpoest:2025gui}, but the availability of the electromagnetic energy measurement and $X_\mathrm{max}$ of each shower will enable more thorough tests with IceCube-Gen2.
Moreover, the production of prompt muons can be studied at a different level than with IceCube.
Although IceCube has measured muons up to the PeV energy range~\cite{IceCube:2015wro}, these events typically lack the measurement of the parent air shower because the shower axis misses IceTop due to its limited size.
With the 30-fold increase of the aperture for surface-deep coincident events, this will fundamentally change with IceCube-Gen2.
Because many prompt muons are expected from proton showers around a PeV of energy and would take a significant fraction of the total shower energy~\cite{Fedynitch:2018cbl}, measuring the parent showers with full efficiency requires a low detection threshold for the surface array.
Next to the veto function, the study of prompt muons has therefore been another criterion to design the surface array to provide full efficiency for protons above $0.5\,$PeV.

The simultaneous measurement of $X_\mathrm{max}$ and the high-energy muon content will also enhance the accuracy for the mass composition.
Moreover, the increased aperture of the surface-deep coincidences in combination with the per-event mass sensitivity will significantly enhance the field-of-view to search for mass dependence of the cosmic-ray anisotropies.
The increase in statistics alone will enable to reduce the gap of statistically significant anisotropy measurements: if the size of the dipole anisotropy would indeed reach the percent-level above $10$'s of PeV, as indicated by initial measurements which are not statistically significant, then IceCube-Gen2 could provide a definitive measurement of that anisotropy up to about the energy range of the second knee (see Fig.~\ref{fig_anisotropy}).
The measurement of expected differences of that anisotropy between mass groups can then provide additional hints to the origin of the highest energy Galactic cosmic rays and the Galactic-to-extragalactic transition.

\begin{figure}[t]
\begin{center}
    \vspace{-0.7cm}
    \includegraphics[width=0.99\linewidth]{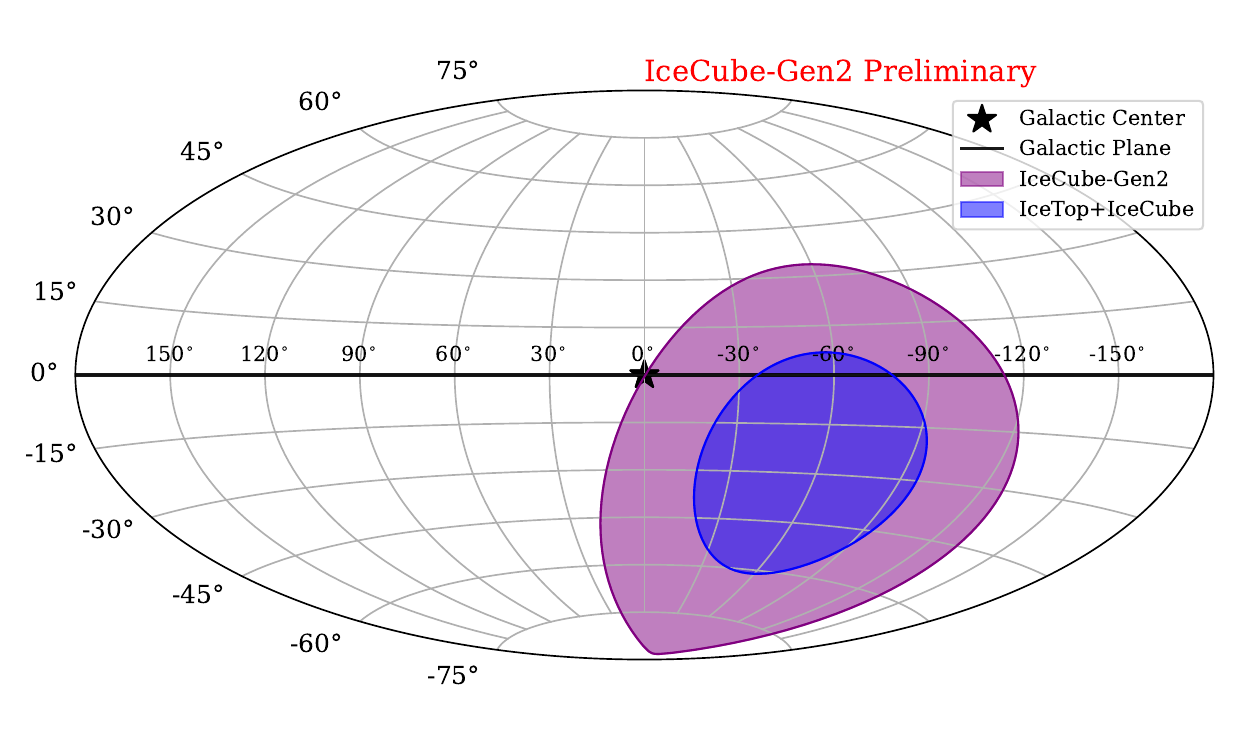}
    \caption{Estimated field-of-view (FOV) of IceCube-Gen2 for gamma rays at several PeV, calculated by the maximum zenith angle for the air-shower axis to intersect both the surface and optical arrays, assuming $\theta_\mathrm{max}^\mathrm{IceCube}=34^\circ$ and $\theta_\mathrm{max}^\mathrm{Gen2}=60.8^\circ$. The exact FOV may be smaller or larger, as it will depend on the criteria of particular analyses and because the detection efficiency depends on the zenith angle (cf.~Fig.\ref{fig_threshold}).
    }
    \label{fig_FOV}
\end{center}
\end{figure}

\section{Searching for PeV Gamma Rays}
As the detection efficiency of the surface array will be similar for gamma rays as for protons, IceCube-Gen2 can also be used as a detector for gamma-rays in the sub-PeV to PeV energy range.
Photon-induced air showers feature significantly less muons than hadronic air showers. 
In particular, almost all proton-induced air showers above $10^{14.5}\,$eV contain muons of energy high enough to be detectable in the optical array, while most gamma-ray induced air showers up to a few PeV of energy do not.
IceCube has already demonstrated that the deep array can be used as a veto to suppress hadronic cosmic-rays by several orders of magnitude relative to photon-induced air showers~\cite{IceCube:2019scr,BontempoThesis}.
Although methods for gamma-hadron separation have not yet been studied specifically for IceCube-Gen2, we therefore expect that it will also serve as a gamma-ray observatory above a few $100$'s of TeV.

One of the major drawbacks of IceCube's searches for photon sources is the limited field-of-view (FOV). 
In the past years, LHAASO, an air-shower array in the Northern hemisphere, has detected several sources of PeV gamma rays in the Galactic Plane~\cite{LHAASO:2021gok}. 
However, IceCube's FOV contains only a small part of the Galactic Plane and with limited exposure, as the shower axis needs to intersect both the surface and optical array for effective gamma-hadron separation.
The increased range of zenith angles for which those types of intersecting events can be observed with IceCube-Gen2 translates into a larger FOV (Fig.~\ref{fig_FOV}), containing significant parts of the Galactic Plane, complementary to LHAASO's FOV. 
Moreover, the aperture for IceCube's existing FOV will be significantly increased: by about the factor 8 of increased surface area for vertical events, and even more for inclined events.
Therefore, IceCube-Gen2 will play an important role to determine which gamma-ray sources in the Southern hemisphere extend to PeV energies.
Hence, together with the observation of Galactic neutrinos~\cite{IceCube:2023ame}, also the detection of photons could directly reveal sources of the highest energy Galactic cosmic rays.

\section{Conclusion}
IceCube-Gen2 with its surface array fulfills a need for higher accuracy of air-shower measurements~\cite{Coleman:2022abf} and will contribute to advancing the particle physics and astrophysics of cosmic rays in the energy range up to a few EeV.
In particular, due to the drastically enhanced statistics and additional radio measurement of the electromagnetic shower energy, IceCube-Gen2 will significantly enhance IceCube's capabilities.
On the particle physics side, the combined measurement of the air shower at the surface and the high-energy muon content will provide unique possibilities to study hadronic interactions in air showers. 
The same increases in statistics and accuracy for the shower measurements will also advance the astrophysics of cosmic rays, studying the origin of the highest energy Galactic cosmic rays and the transition to extragalactic cosmic rays.

\bibliographystyle{ICRC}
\bibliography{bibliography}

\clearpage

\section*{Full Author List: IceCube-Gen2 Collaboration}

\scriptsize
\noindent
R. Abbasi$^{16}$,
M. Ackermann$^{76}$,
J. Adams$^{21}$,
S. K. Agarwalla$^{46,\: {\rm a}}$,
J. A. Aguilar$^{10}$,
M. Ahlers$^{25}$,
J.M. Alameddine$^{26}$,
S. Ali$^{39}$,
N. M. Amin$^{52}$,
K. Andeen$^{49}$
% ...
\\
\\
% ...
$^{10}$ Universit{\'e} Libre de Bruxelles, Science Faculty CP230, B-1050 Brussels, Belgium \\
% ...
$^{16}$ Department of Physics, Loyola University Chicago, Chicago, IL 60660, USA \\
% ...
$^{21}$ Dept. of Physics and Astronomy, University of Canterbury, Private Bag 4800, Christchurch, New Zealand \\
% ...
$^{25}$ Niels Bohr Institute, University of Copenhagen, DK-2100 Copenhagen, Denmark \\
$^{26}$ Dept. of Physics, TU Dortmund University, D-44221 Dortmund, Germany \\
% ...
$^{39}$ Dept. of Physics and Astronomy, University of Kansas, Lawrence, KS 66045, USA \\
% ...
$^{46}$ Dept. of Physics and Wisconsin IceCube Particle Astrophysics Center, University of Wisconsin{\textemdash}Madison, Madison, WI 53706, USA \\
% ...
$^{49}$ Department of Physics, Marquette University, Milwaukee, WI 53201, USA \\
% ...
$^{52}$ Bartol Research Institute and Dept. of Physics and Astronomy, University of Delaware, Newark, DE 19716, USA \\
% ...
$^{76}$ Deutsches Elektronen-Synchrotron DESY, Platanenallee 6, D-15738 Zeuthen, Germany \\
$^{\rm a}$ also at Institute of Physics, Sachivalaya Marg, Sainik School Post, Bhubaneswar 751005, India \\
% ...

\subsection*{Acknowledgments}

\noindent
The authors gratefully acknowledge the support from the following agencies and institutions:
USA {\textendash} U.S. National Science Foundation-Office of Polar Programs,
U.S. National Science Foundation-Physics Division,
U.S. National Science Foundation-EPSCoR,
U.S. National Science Foundation-Office of Advanced Cyberinfrastructure,
Wisconsin Alumni Research Foundation,
Center for High Throughput Computing (CHTC) at the University of Wisconsin{\textendash}Madison,
Open Science Grid (OSG),
Partnership to Advance Throughput Computing (PATh),
Advanced Cyberinfrastructure Coordination Ecosystem: Services {\&} Support (ACCESS),
Frontera and Ranch computing project at the Texas Advanced Computing Center,
U.S. Department of Energy-National Energy Research Scientific Computing Center,
Particle astrophysics research computing center at the University of Maryland,
Institute for Cyber-Enabled Research at Michigan State University,
Astroparticle physics computational facility at Marquette University,
NVIDIA Corporation,
and Google Cloud Platform;
Belgium {\textendash} Funds for Scientific Research (FRS-FNRS and FWO),
FWO Odysseus and Big Science programmes,
and Belgian Federal Science Policy Office (Belspo);
Germany {\textendash} Bundesministerium f{\"u}r Forschung, Technologie und Raumfahrt (BMFTR),
Deutsche Forschungsgemeinschaft (DFG),
Helmholtz Alliance for Astroparticle Physics (HAP),
Initiative and Networking Fund of the Helmholtz Association,
Deutsches Elektronen Synchrotron (DESY),
and High Performance Computing cluster of the RWTH Aachen;
Sweden {\textendash} Swedish Research Council,
Swedish Polar Research Secretariat,
Swedish National Infrastructure for Computing (SNIC),
and Knut and Alice Wallenberg Foundation;
European Union {\textendash} EGI Advanced Computing for research;
Australia {\textendash} Australian Research Council;
Canada {\textendash} Natural Sciences and Engineering Research Council of Canada,
Calcul Qu{\'e}bec, Compute Ontario, Canada Foundation for Innovation, WestGrid, and Digital Research Alliance of Canada;
Denmark {\textendash} Villum Fonden, Carlsberg Foundation, and European Commission;
New Zealand {\textendash} Marsden Fund;
Japan {\textendash} Japan Society for Promotion of Science (JSPS)
and Institute for Global Prominent Research (IGPR) of Chiba University;
Korea {\textendash} National Research Foundation of Korea (NRF);
Switzerland {\textendash} Swiss National Science Foundation (SNSF).

\end{document}